\documentclass[11pt,twoside]{article}

\usepackage{asp2004}
\usepackage{epsf}
\usepackage{psfig}
\usepackage{graphicx}
\usepackage{lscape}

\usepackage[]{natbib}

\begin{document}
\title{The $N$-Body Approach to Disk Galaxy Evolution}   
\author{Victor P. Debattista}   
\affil{Brooks Fellow, Astronomy Department, University of Washington,
Box 351580, Seattle, WA 98195, USA}    

\begin{abstract} 
I review recent progress from $N$-body simulations in our
understanding of the secular evolution of isolated disk galaxies.  I
describe some of the recent controversies in the field which have been
commonly attributed to numerics.  The numerical methods used are
widely used in computational astronomy and the problems encountered
are therefore of wider interest.
\end{abstract}



\section{Introduction}
\label{sec:intro}

The fragility of disk galaxies suggests that a significant fraction of
their history was spent experiencing mild external perturbations.
Thus their evolution is likely to have been driven partly by internal
processes.  Dissipational gas physics probably plays a prominent role
in this evolution but the effects of collisionless evolution very near
equilibrium are also significant.  Collisionless secular evolution is
driven largely by non-axisymmetric structures, especially by bars,
which represent large departures from axisymmetry, and are thus
efficient agents of mass, energy and angular momentum redistribution.
Moreover bars occur in over $50\%$ of disk galaxies \citep{knapen_99,
esk_etal_00} and simulations starting from the 1970s
\citep{mil_etal_70, hohl_71} have shown that they form readily via
global instabilities \citep{kalnaj_72,toomre_81}.  Here I review
recent progress in understanding the secular evolution of isolated
disk galaxies.


\section{Numerical Methods}
\label{sec:numerics}

At present there is considerable debate on the role numerics play in
the evolution of $N$-body simulations.  This review therefore treats
collisionless isolated galaxy simulations as the necessary
prerequisites for confidence in any other $N$-body study of galaxy
formation and evolution.  The physics of isolated collisionless
systems consist only of Newton's law of gravity and Newtonian
mechanics, both of which are well understood.  Other than initial
conditions, the main difficulty lies in computing the gravitational
field of a given mass distribution accurately and efficiently.  Any
gravodynamical code must include at least this much, including those
used for studying planet and cosmological structure formation.

Many algorithms have been devised for calculating gravitational fields
of isolated galaxies.  The simplest is direct particle-particle
interactions \citep[e.g.][]{aarset_63}, but this scales as ${\cal
O}(N^2)$, $N$ being the number of particles, making it impractical for
other than special applications.
Particle-Mesh (PM/grid) codes \citep{mil_pre_68,miller_76} solve the
potential on a grid by binning the mass distribution.  Several
geometries are possible, including cartesian, cylindrical and
spherical.  Grid codes are very efficient, scaling largely with the
number of grid cells, $N_g$, as $\sim {\cal O}(N_g)$
\citep{sellwo_97}.  Their main drawbacks are their inflexible geometry
and their trade-off between volume and spatial resolution, but hybrid
codes can alleviate many of these problems \citep[e.g.][]{fux_99,
sellwo_03}.
Adaptive Mesh Refinement (AMR) codes \citep[e.g.][]{bry_nor_95,
kra_etal_97}, on the other hand, let the grid evolve dynamically to
increase the resolution in dense regions.
Tree codes \citep[e.g.][]{bar_hut_86,hernqu_87} instead group
particles by location, computing direct forces for nearby particles
and a few multipoles for the more distant ones.  These codes can be
vectorized efficiently and scale as ${\cal O}(N\log N)$
\citep{dubins_96, stadel_phd, spr_etal_01} while \citet{dehnen_00}
presented an ${\cal O}(N)$ extension.  Tree codes have been widely
implemented and several are publicly available.
Self-Consistent Field (SCF) codes \citep[e.g.][]{clutto_72,
her_ost_92, ear_sel_95} expand the density in a set of orthogonal
basis functions, from which the potential can then be computed.  SCF
codes scale as ${\cal O}(N)$ and do not require softening because
truncating the expansion is enough to suppress small scale noise.
However they require that the basis set be chosen with care or that a
large number of basis functions be included (which re-introduces small
scale noise).  \citet{weinbe_96} solves this problem by making the
basis set adaptive.

\subsection{Code testing}

The first requirement of any $N$-body code used to study galaxy
evolution is that it be collisionless.  Relaxation rates of conserved
quantities are useful diagnostics for this purpose
\citep[e.g.][]{hohl_73, her_bar_90, her_ost_92, val_kly_03}.

Testing the gravodynamical part of a code requires more effort.
Comparison with the limited number of exact analytic results known
provide stringent code tests; such systems (in 2-D) include the
Kalnajs disk \citep{kalnaj_72}, the Mestel disk \citep{zang_phd}, the
isochrone disk \citep{kalnaj_78} and the power-law disks
\citep{eva_rea_98a, eva_rea_98b}.
Several code tests using these predictions have been carried out.
\citet{ear_sel_95} compared the eigenfrequencies of instabilities in
isochrone disks with simulations using an SCF and a PM method.  They
were able to reproduce the predicted values to within $5\%$ with just
$15K$ particles using the SCF method whereas softening caused the PM
code to never quite converge to the predicted value.
Meanwhile, \citet{sel_eva_01} presented $N$-body examples of the
power-law disks including reproducing a challenging case of a
perfectly stable disk.  They reported that aliasing rendered SCF codes
ill-suited to these disks.
Instabilities of spherical systems have also been used for code
testing.  For example \citet{weinbe_96} tested his adaptive SCF code
on the instabilities of spherical generalized polytropes investigated
by \citet{barnes_etal_86} and found generally good agreement.
Analytic solutions of non-equilibrium evolution, which include the
solution of 1-D plane wave collapse \citep{zeldov_70} and of spherical
infall in an expanding universe \citep{fil_gol_84,bertsc_85} have also
be used to test $N$-body codes \citep{kra_etal_97,dave_etal_97}.

Different codes can be tested also by direct comparison.
\citet{ina_etal_84} compared bar formation between an $N$-body model
and a direct numerical integration of the CBE and Poisson equation
\citep{nis_etal_81,wat_etal_81}.  The two simulations matched each
other to better than $2\%$ in bar amplitude well into the non-linear
regime and eventual discrepancies were due to the inability of the CBE
code to handle large gradients in the distribution function (DF).


\section{Cusp evolution}
\label{sec:cusps}

A prediction of cold dark matter (CDM) cosmology is that dark matter
halos are cusped, with densities $\rho \sim r^{-\beta}$ at small radii
and $1 \leq \beta \leq 1.5$ \citep{nav_etal_97_nfw, moo_etal_98,
jin_sut_00, pow_etal_03}.  Several arguments have been advanced
against the presence of such cusps in real galaxies, including
detailed rotation curve fits of dwarf and low surface brightness
galaxies \citep{bla_etal_01, deblok_etal_01, mat_gal_02}, bar gas
flows \citep{wei_etal_01b, per_etal_04} and pattern speeds \citep[see
\S \ref{sec:dynfric} below]{deb_sel_98, deb_sel_00} and the
microlensing optical depth and gas dynamics in the Milky Way
\citep{bin_etal_00, bis_etal_03}.

Several ways to erase cusps have been considered including new dark
matter physics \citep{spe_ste_00, peeble_00, goodma_00, kap_etal_00,
cen_01}, feedback from star formation \citep{nav_etal_96, gne_zha_02},
or an initial power spectrum with decreased power on small scales
\citep{hog_dal_00, col_etal_00, bod_etal_01}.  \citet{bin_etal_01}
suggested that bars in young galaxies were able to torque up cusps and
expel them, while \citet{her_wei_92} had found that such torques can
reduce the density of a spheroid by a factor of $\sim 100$ out to
$\sim 0.3 a_B$ (where $a_B$ is the bar's semi-major axis).
\citet[hereafter WK02]{wei_kat_02}, following \citet{her_wei_92},
presented perturbation theory calculations and simulations of imposed
rigidly-rotating non-slowing (IRRNS) bars to argue that cusps are
erased in a few bar rotations ($\sim 10^8$ years).  Their bars were
large ($\sim 10$ kpc), but they argued for a scenario in which large
primordial bars destroy cusps while the current generation of smaller
bars formed later.  Large reductions of dark halo densities were not
seen in self-consistent cuspy halo simulations
\citep[e.g.][]{one_dub_03, val_kly_03}; WK02 suggested that this was
because the resonant dynamics responsible for cusp removal are very
sensitive to numerical noise.  They found that they needed $N > 10^6$
for their SCF code and argued that even larger $N$ would be needed for
grid, tree or direct codes.

The scenario of WK02 makes three main claims: (1) that bars cause a
decrease in cusp density and can destroy cusps if sufficiently large
and strong (2) that such bars formed via interactions at high redshift
and (3) that $N$ needs to be large in order that the relevant phase
space is adequately covered and that orbit diffusion does not destroy
the resonant dynamics causing cusp destruction.

\subsection{When do bars destroy cusps?}

\citet[][hereafter S03]{sellwo_03} presented simulations of the same
IRRNS bar as was used by WK02 and found cusp destruction occurred in a
runaway process.  The inclusion of the $l=1$ spherical harmonic terms
in the potential had a large effect on the evolution; in their
absence, cusp removal took $\sim 5-6$ times longer than when they were
present.  For the same IRRNS bar, \citet[][hereafter MD05]{mcm_deh_05}
found that the cusp moves off-center by as much as $30\%$ of $a_B$.
This centering instability gives the appearance that the cusp has been
erased when halo density is measured by spherically averaging about
the origin which, they argued, caused WK02 to over-estimate a bar's
ability to erase a cusp.  Evidence for lopsidedness in the simulation
of WK02 can be inferred from its asymmetric bar-induced halo wake
(their figure 2).  When MD05 suppressed this purely numerical
instability they still found that the cusp is destroyed, although
after a much longer time.  The possibility that an offset cusp is
being confused for cusp destruction was investigated by
\citet[hereafter HB05]{hol_etal_05}.  Their self-consistent
simulations formed bars in a disk of particles, thus damping any
centering instability.  While confirming that some of the evolution
seen by WK02 was due to the centering instability, they found that
when odd $l$ terms were excluded in their simulations that the damage
to the cusp was not significantly diminished.

Thus these idealized bars at least are able to destroy cusps.  What
about more realistic bars?  S03 showed that when, instead of assuming
a fixed pattern speed, $\Omega_{\rm p}$, he allowed it to decrease
such that total angular momentum is conserved assuming a constant
moment of inertia (an IRRS bar), that the cusp was damaged
significantly less.  This happens because an IRRNS bar must transfer
more angular momentum to the halo than it can plausibly have in order
to destroy the cusp.  Shorter, more realistically sized bars were also
ineffective bar destroyers.  Similarly MD05 found that an IRRS bar is
unable to destroy a cusp in a Hubble time.  S03 also presented
self-consistent simulations in which a disk of particles was grown
inside a halo.  In these cases, besides forming smaller bars which
slow, cusp erasure was also inhibited by the increase in halo density
as the disk grew and again as the central density of the disk
increased because of angular momentum transport outwards.  HB05
instead found that the large relative angular momentum gained by the
inner halo in their self-consistent simulations flattened the cusp,
but only out to 900 pc scaled to the Milky Way.

\subsection{Do bars get as large as needed?}

Thus all self-consistent simulations find that bars of sizes typical
of those observed are unable to remove cusps on scales of several kpc.
Do bars ever get to be large enough to do so?  Bars that form via disk
instabilities \citep{toomre_81} generally extend to roughly the radius
of the rotation curve turnover.  However, externally triggered bars
may be substantially larger.  HB05 showed an example of such a bar;
scaled to the Milky Way, external triggering produced a bar of $a_B =
12$ kpc, as opposed to $\sim 4.5$ kpc via the bar instability.
Whether the required triggering actually occurs is a question for
hierarchical simulations while direct observations at high redshift
should establish whether bars get to be as large as 10 kpc.

\subsection{Is the evolution in simulations compromised by too small $N$?}

\citet{wei_kat_05} provided several reasons for the need of large $N$.
They focused especially on the inner Lindblad resonance (ILR) which
always extends down to the cusp.  One possible problem they identified
is two-body scattering which leads to particles executing a random
walk and therefore lingering in resonance for significantly less than
they would otherwise.  In the terminology of \citet{tre_wei_84},
scattering causes particles to traverse the ILR in the fast rather
than the slow regime.  Since resonant torques scale as $m_P^2$ in the
fast regime and as $\sqrt{m_P}$ in the slow regime (where $m_P$ is the
fractional mass of the perturber) two-body scattering causes the
friction to be substantially reduced.  For a typical bar, they argued
that $N > 10^8$ within the virial radius is required.  A second
limiting factor they identified is phase space coverage.  Whether a
resonant particle gains or loses angular momentum depends on its
phase.  If the phase space of the resonance is not sufficiently
sampled by particles then the correct ensemble average is not
attained.

HB05 found no difference between their tidally perturbed simulations
with $N = 5.5M$ and $N = 11M$, while $N = 1.1M$ resulted in a weaker
bar and less friction than in the other two cases.  This, they argued,
proved that the $N = 1.1M$ simulation suffered from too much orbital
diffusion to follow resonances correctly.  However, \citet{sellwo_05}
argued plausibly that this $N$ dependence was due to the bar being
weaker, and not because friction depends on $N$.  The weaker bar in
the $N=1.1M$ simulation was caused by the unavoidably larger $m=2$
{\it seed} amplitude in the initial disk which, once swing-amplified,
interfered destructively with the distortion induced by the externally
applied tidal field.

The simulations of S03 (using both a PM and an SCF code) and MD05
(using a tree code) are instructive because they studied the same
physical system: a \citet{hernqu_90} halo with a IRRNS bar of $a_B =
0.7 r_s$ ($r_s$ being the halo scale radius).  Both studies found that
the onset of the runaway cusp destruction depended on $N$ and occurs
earlier with {\it decreasing} $N$.  S03 found good agreement between
evolution using the PM code and the SCF code; additionally he showed
that using only $l=0$ and 2 terms was sufficient for the evolution,
with little change when larger even values of $l$ were included.  On
the other hand, MD05 forced symmetry about the origin to suppress the
centering instability.  Thus the two studies were very similar; but
while their results are in qualitative agreement, there are surprising
quantitative differences.  S03 found that the onset of the runaway
appears to be converging by $N = 3M$ at $\sim 130$ bar rotations (his
figure 2b) but for the same $N$ MD05 saw no evidence for convergence
with runaway at $\sim 210$ bar rotations (their figure 5b).  Yet both
agreed that the radius containing $1\%$ of the total mass is driven
from $\sim 0.1 r_s$ to $\sim 0.5 r_s$.  Unless the difference is due
to the initial conditions in some worrisome way, noise would seem to
responsible for these differences.  A more careful comparison between
these different simulations seems particularly worthwhile.

\citet{sel_deb_05} report an entirely different test of whether
scattering overwhelms bar evolution.  They presented simulations
(using a PM code) in which they perturbed the system into a metastable
state in which $\Omega_{\rm p}$ was nearly constant (more on this in
\S \ref{sec:dynfric}).  Their metastable state resulted because
$\Omega_{\rm p}$ was driven up such that the principal resonances are
trapped in {\it shallow} local minima of the DF.  Systems persisted in
this metastable state for $\sim 5$ Gyr, which would not have been
possible if orbit scattering had been strong.

Unlike two-body scattering, phase space coverage can be investigated
by means of test particle simulations, where the self-gravitating halo
response is not included.  Using such experiments, \citet{sellwo_05}
found that $N \sim 1M$ was sufficient for the $\Omega_{\rm p}$
evolution of a low mass bar $M_{\rm bar} = 0.005 M_{\rm halo}$ to
converge, and a factor of $\sim 100$ less particles were needed for
$M_{\rm bar} = 0.02 M_{\rm halo}$.  When self-gravity was introduced
(thus adding scattering), $1M$ particles were needed for the $M_{\rm
bar} = 0.02 M_{\rm halo}$ case.  As in S03, he argued that WK02 needed
large $N$ because of the extremely difficult nature of their
experiments with a fixed-amplitude bar rotating at a fixed
$\Omega_{\rm p}$.  When $\Omega_{\rm p}$ is a function of time, the
resonances are broadened to orbits near resonance, which makes phase
space coverage a less stringent constraint on $N$.  Weak bars in
particular may be prone to such difficulties, but it is strong bars
which are most interesting.

In summary, there may be some evidence that noise is compromising some
if not all of the evolution in some cases, in ways not yet fully
understood.  But it is clear that only quite large bars can erase
cusps to scales larger than 1 kpc.


\section{Evolution of pattern speeds}
\label{sec:dynfric}

The pattern speed of a bar is usually parametrized by the ratio ${\cal
R} \equiv D_L/a_B$, where $D_L$ is the corotation radius, at which the
gravitational and centrifugal forces cancel.  A self-consistent bar
must have ${\cal R} \geq 1$ \citep{contop_81}.  A bar is termed fast
when $1.0 \leq {\cal R} \leq 1.4$ and slow otherwise.  This definition
does not distinguish fast from slow in terms of $\Omega_{\rm p}$
alone: a bar in a galaxy with rotation velocity of $200$ $\mathrm
{km}~\mathrm{s}^{-1}$ is slow at $\Omega_{\rm p} = 100$ $\mathrm
{km}~\mathrm{s}^{-1}~ \mathrm{kpc}^{-1}$ if $a_B = 1$ kpc, but fast at
$\Omega_{\rm p} = 20$ $\mathrm {km}~\mathrm{s}^{-1}~
\mathrm{kpc}^{-1}$ if $a_B = 10$ kpc.  Observational evidence points
to fast bars, both in early-type barred galaxies \citep{mer_kui_95,
gersse_etal_99, deb_cor_agu_02, agu_etal_03, gersse_etal_03,
deb_wil_04} and in late-types \citep{lin_etal_96, lin_kri_96,
wei_etal_01b, per_etal_04}.

\citet{tre_wei_84} developed the perturbation theory of dynamical
friction for perturbers in spheroidal systems, showing that friction
arises near resonances, when $m \Omega_{\rm p} = k \Omega_r + l \Omega_\phi$,
where $\Omega_r$ and $\Omega_\phi$ are the radial and angular
frequencies, respectively \citep[see][for a time-dependent
treatment]{weinbe_04}.  \citet{weinbe_85} applied this theory to a bar
rotating in a dark halo and found that the bar is braked such that
$D_L \gg a_B$ unless (1) angular momentum is added to the bar, (2)
the bar is weak or (3) the halo has low mass.

The transfer of angular momentum from disk to spheroid (bulge or halo)
was reported in several early simulations \citep{sellwo_80,
lit_car_91, her_wei_92}.  Fully self-consistent, 3-D simulations of
bar-unstable disks embedded in dark halos were presented by
\citet[][hereafter DS00]{deb_sel_98, deb_sel_00} who found that disks
needed to be the dominant mass component if the bars which form were
to remain at ${\cal R} < 1.4$.  Bar slowdown has also been found in other
simulations since then \citep{one_dub_03, hol_etal_05, ber_etal_05,
mar_etal_05}.  \citet{one_dub_03} included a comparison with DS00 and
found generally good agreement.

The constraint of DS00 severely limits CDM cusps to be present in dark
halos.  \citet[][hereafter VK03]{val_kly_03} presented simulations of
a CDM Milky Way, with halo concentration $c \simeq 15$.  These
simulations were evolved on an AMR code with a maximum refinement
corresponding to a resolution of 20-40 pc.  They found that the bar
which formed remained fast for $\sim 4-5$ Gyr, thereafter slowing to
${\cal R} = 1.7$.  They concluded that the slow bars in the
simulations of DS00 were an artifact of low resolution.  They also
argued that their bars remained modest in length (1-2 scale-lengths)
but became too long in lower resolution simulations, which they
concluded again proved the limitations of low resolution simulations.
\citet{sel_deb_05} repeated one of the simulations of VK03 using the
same initial conditions but evolved on a hybrid grid code
\citep{sellwo_03} with a fixed softening comparable to that of VK03.
They found instead that the bar reached ${\cal R} = 2$ within 4 Gyr.
They accounted for the result of VK03 by noting that as the bar
formed, the central density increased (see \S \ref{sec:structure}).
Thus in an AMR code there is a tendency for the spatial resolution to
increase; the disk in the model of VK03 being rather thin, this led to
enhanced forces and a corresponding artificial increase in
$\Omega_{\rm p}$.  Once this happened, the bar found itself with its
principal resonances at local phase space minima previously generated
by the forming bar.  Although angular momentum is exchanged at
resonances, the sign of the torque depends on the phase space gradient
of the DF at the resonances; in the absence of a gradient no friction
is possible.  Thus the numerics induced a metastable state.  This
state proved fragile to realistic perturbations and is not likely to
last long in nature, but in the quieter environment of an isolated
$N$-body system it can persist for many Gyrs, enough to fully account
for the behavior found by VK03.

In a series of papers, Athanassoula presented several arguments
against the conclusion of DS00.  In \citet{athana_03} she compared the
evolution of two halo-dominated systems, MQ2 and MHH2, with nearly
identical disk and halo rotation curves.  However, their velocity
dispersions were different, being larger in MHH2 because its halo
extended to larger radii.  As a result $\Omega_{\rm p}$ decreased
significantly in MQ2 but hardly at all in MHH2.  She concluded that
this ``argues against a link between relative halo content and bar
slowdown...''  This claim however is contradicted by her own
simulations.  The confusion arises from her relying on the change in
$\Omega_{\rm p}$ to constrain the halo.  Not only is this
unobservable, but if we compute ${\cal R}$ for her models, we find
$1.4 < {\cal R} \leq 1.7$ for run MQ2 and ${\cal R} \simeq 3$ for
MHH2, {\it i.e.} both these halo-dominated systems are slower than
observed.  Far from being in disagreement with the results of DS00,
her simulations support them.  She also showed that weak bars are less
able to drive angular momentum exchanges and may remain fast; since
observational measurements have only been obtained for strong bars,
this is not worrisome.

\citet{ath_mis_02} argued that bar lengths are difficult to measure in
simulations making a comparison with corotation difficult.  Bar
lengths are certainly not always easy to measure, but it is still
possible to define values which straddle the real value; for example
DS00 had cases in which the uncertainty in $a_B$ was as large as
$\Delta a_B/\overline{a_B} \sim 0.3$, not substantially different than
in observations \citep{debatt_03}.  But these are not Gaussian errors
and the probability of $a_B$ falling outside the given range is
practically zero.  Moreover, the same problem afflicts observations;
therefore measurement uncertainties in $a_B$ are a nuisance but not a
repudiation of the constraint.

\citet{athana_02} showed that loss of angular momentum from the bar
leads to a growing bar, as first suggested by DS00.  VK03 pointed out
that bars become excessively long in the presence of strong friction.
Even though angular momentum redistribution leads to larger disk
scale-lengths, bars extending $\ga 10$ kpc are not common
\citep{erwin_05}.  Strong bar growth does not seem to have occurred
through the history of the current generation of bars and presumably,
neither has strong friction.

Thus to date no well-resolved simulation has provided a valid
counter-example to the claim by \citet{deb_sel_98} that dense halos
cannot support fast strong bars.  If anything, recent simulations have
lent support to it.


\section{The secular evolution of disk densities}
\label{sec:structure}

The excellent recent review by \citet{kor_ken_04} presents the
observational evidence for pseudo-bulge formation via secular
evolution and discusses some of the older $N$-body results in that
field.  Here I review recent developments not covered by those
authors.

\subsection{Disk Profile Evolution}

Bar formation is accompanied by a rearrangement of disk material as
first shown by \citet{hohl_71}.  Generally a nearly double-exponential
profile develops, with a smaller central scale-length than the initial
and a larger one further out.  In this respect these profiles resemble
bulge+disk profiles and comparisons with observations show that these
profiles are reasonable approximations to observed profiles
\citep{deb_etal_04, avi_etal_05}.  \citet{deb_etal_05b} showed that
the degree by which the profile changes depends on the initial disk
temperature $Q$.  In hot disks ($Q \sim 2$) little angular momentum
needs to be shed and the azimuthally-averaged density profile is
practically unchanged.  Thus the distribution of disk scale-lengths
depends not only on the initial angular momentum of the baryons (and
presumably that of the dark halo) but also the disk temperature.
Angular momentum redistribution continues also after the bar forms,
especially to the halo.  This leads to a further increase in the
central density of the disk even when the evolution is collisionless.
This may be sufficient to render an initially halo-dominated system
into one dominated by baryons in the inner parts \citep{deb_sel_00,
val_kly_03}.

The angular momentum lost by the bar as it forms may be transported
out to large radii via a resonant coupling between the bar and spirals
\citep{deb_etal_05b} of the kind found by \citet{mas_tag_97} and
\citet{rau_sal_99}.  \citet{deb_etal_05b} show that this transport
leads to breaks in the density distribution which, when viewed
edge-on, are indistinguishable from those observed in real galaxies
\citep{pohlen_phd}.

\subsection{Vertical Evolution: Peanut-Shaped Bulges}

$N$-body simulations of the vertical evolution of disk galaxies have
concentrated on box- or peanut- (B/P-) shaped bulges which are present
in some $45\%$ of edge-on galaxies \citep{lut_etal_00}.  Simulations
have shown that these form via the secular evolution of bars
\citep{com_san_81}, either through resonant scattering or through
bending (aka buckling) instabilities \citep{pfenni_84, com_etal_90,
pfe_fri_91, rah_etal_91}.  Although it is often thought that a peanut
requires that a bulge was built by secular processes, simulations show
that peanuts can also form when the initial conditions include a
bulge, as would happen if bulges form through mergers at high redshift
\citep{ath_mis_02, deb_etal_05a}.

Observations seeking to establish the connection between B/P-shaped
bulges and bars \citep{kui_mer_95, mer_kui_99, bur_fre_99,
chu_bur_04}, by looking for evidence of bars in edge-on B/P-bulged
systems, have benefited from comparisons with the edge-on stellar
velocity distributions of $N$-body bars \citep{bur_ath_99,
bur_ath_05}.  \citet{bur_ath_05} characterized the signature of an
edge-on bar as having (1) a Freeman type II profile, (2) a rotation
curve with a local maximum interior to its flat part (3) a broad
velocity dispersion profile with a plateau at intermediate radii (4) a
correlation between velocity and the third-order Gauss-Hermite moment
$h_3$ \citep{gerhar_93, vdm_fra_93}.  A diagnostic of B/P-shaped
bulges in face-on galaxies was developed, and tested on high mass and
force resolution simulations, by \citet{deb_etal_05a}.  Vertical
velocity dispersions constitute a poor diagnostic because they depend
on the local surface density.  Instead, their diagnostic is based on
the fact that peanut shapes are associated with a flat density
distribution in the vertical direction.  The kinematic signature
corresponding to such a distribution is a minimum in the fourth-order
Gauss-Hermite moment $h_4$.

The buckling instability itself has also been studied with
simulations.  \citet{deb_etal_05b} showed that an otherwise
vertically-stable bar is destabilized when it slows and grows.
Moreover, the instability can occur more than once for a given bar
\citep{mar_etal_05}.  Finally, after \citet{rah_etal_91} showed that
buckling weakens bars, it has often been assumed that bars are
destroyed by buckling.  \citet{deb_etal_05b} presented a series of
high force and mass resolution simulations demonstrating that this is
not the case.

\subsection{Spirals}

There is broad agreement that spirals constitute density waves
\citep{lin_shu_64}.  At least three different dynamical mechanisms
have been proposed for exciting them: swing-amplification
\citep{toomre_81}, long-lived modes \citep{ber_etal_89a, ber_etal_89b}
and recurrent instabilities seeded by features in the angular momentum
distribution \citep{sel_lin_89, sel_kah_91, sellwo_00b}.  All three
are still viable and not much new $N$-body results have been obtained
in recent years, but \citet{sel_bin_02} used simulations to
demonstrate that spirals cause a considerable radial shuffling of mass
at nearly fixed angular momentum distribution.  This happens at
corotation and is not accompanied by substantial heating --- stars on
nearly circular orbits can be scattered onto other nearly circular
orbits.  For example, they estimate that a star born at the solar
radius can be scattered nearly uniformly within $\Delta R = \pm 4$
kpc.  Thus the idea of a Galactic habitable zone becomes somewhat
suspect, as does the need for infall to maintain the metallicity
distribution observed at the solar circle.


\section{Bar destruction by CMCs}
\label{sec:cmcs}

Central massive concentrations (CMCs), whether supermassive black
holes (hard CMCs) or gas condensations several hundred parsecs in size
(soft CMCs), could destroy bars.  Early studies of this phenomenon
were inspired by the similar work for slowly-rotating triaxial
elliptical galaxies \citep[e.g.][]{ger_bin_85}, where the loss of
triaxiality results from the destruction of box orbits by scattering
off the CMC.  Such scattering may not be efficient in bars, since the
main bar-supporting orbits are loops which avoid the center.
\citet{has_nor_90} and \citet{has_etal_93} argued that, when a CMC in
a barred galaxy grew sufficiently massive, it quickly destroys
bar-supporting orbits by driving an ILR, around which orbits are
unstable, to large radii.  The more centrally concentrated the CMC
grew, the further out was the ILR and therefore the more disruptive it
was.  How massive the CMC needed to be required $N$-body simulations
to establish.  \citet{fri_ben_93} modeled gas and stars with PM+SPH
simulations and found that gas inflows destroyed bars when $2\%$ of
the baryonic mass ended up in a hard CMC.  \citet{nor_etal_96} pursued
collisionless 2-D and 3-D simulations, with CMCs grown by slowly
contracting a massive component.  They needed a CMC of mass $5\%$ of
disk mass, $M_{\rm d}$, to destroy bars.  Then bar destruction was
rapid and led to a bulge-like spheroid.

\citet[][hereafter SS04]{she_sel_04} presented a series of
high-quality simulations including high mass, force and time
resolution and found that bars are more robust than previously
thought.  They varied the growth rates, compactness and mass of the
CMCs and considered both weak and strong bars within which they grew
CMCs at fixed compactness.  A rigid halo with a large core radius was
also included.  They obtained a fast decay while a CMC was growing
followed by a more gradual decay once the CMC reached its full mass.
The time over which the bar was grown proved unimportant, with only
the final mass and compactness mattering.  Hard CMCs cause more damage
than soft ones, needing $4-5\% M_{\rm d}$ and $>10\% M_{\rm d}$,
respectively, to destroy bars.  Their tests showed that time steps
need to be as small as $10^{-4}$ of a dynamical time in order that
more rapid but incorrect bar destruction is avoided.  They interpreted
the two-phase bar destruction as scattering of low energy
bar-supporting $x_1$ orbits during the CMC growth phase and continued
gradual global structural adjustment, which further destroyed high
energy $x_1$ orbits, thereafter.  They predicted that massive halos,
which lead to bar growth \citep{deb_sel_00, athana_03} render bars
even more difficult to destroy; this prediction, as well as the
two-phase bar weakening, was confirmed in live-halo simulations by
\citet{ath_etal_05}.

\citet{bou_com_02} and \citet{bou_etal_05} have argued for a radically
different picture.  They simulated gas accretion and noted that bars
were destroyed and reformed 3 or 4 times over a Hubble time.  The
amount of gas required to destroy bars is not, however, wholly
consistent in these simulations: a system with $\sim 7\%$ {\it total}
gas mass fraction lost its bar within 2 Gyr in \citet{bou_etal_05}
whereas a system with three times more gas maintained its bar in
\citet{bou_com_02}.  Possibly star-formation, included in the later
simulations, somehow quenched the infall onto the center.  SS04 hinted
that the timestep used by \citet{bou_com_02} was too large but
\citet{bou_etal_05} reported that using a timestep $0.125 \times$
their standard value (and close to that advocated by SS04) still led
to recurrent bar destruction.  \citet{bou_etal_05} argued that their
results are correct and that other studies had erred in mimicking gas
accretion by simply growing a massive object because this neglected
the important effects of angular momentum transport from gas to the
bar.  This is, perhaps, consistent with the simulations of
\citet{ber_etal_98} (who were able to destroy bars by gas inflow
leading to a CMC of mass fraction just $1.6\%$) and the earlier ones
of \citet{fri_ben_93}.  On the other hand, the fully live simulations
of \citet{deb_etal_05b} only destroyed bars when soft CMCs reached
$\sim 20\% M_{\rm d}$, in good agreement with SS04.  One possibly
important difference is that the simulations of \citet{bou_etal_05}
included a rigid halo, which prevents it from accepting angular
momentum from the bar, while \citet{deb_etal_05b} had live halos.

\citet{hoz_her_05}, using a 2-D SCF code, also concluded that hard
CMCs can destroy bars with smaller masses, $0.5\%M_{\rm d}$.  SS04
\citep[see also][]{sellwo_02} speculated that low order SCF expansions
may not be able to simultaneously maintain a system axisymmetric near
the center and non-axisymmetric further out.


\section{Multiple patterns}

An emerging field in the past few years has been galaxies with
multiple patterns.  These are challenging to study because traditional
tools such as surfaces-of-section are no longer viable.  $N$-body
simulations, therefore, are vital for studying systems such as bars in
triaxial halos and bars within bars.

\subsection{Bars in triaxial halos}
\label{ssec:3ax}

CDM predicts that dark matter halos are triaxial \citep{bar_efs_87,
frenk_etal_88, dub_car_91, jin_sut_02}.  The condensation of baryons
inside triaxial halos drives them to rounder shapes, but systems do
not generally become wholly axisymmetric \citep{dubins_94, kkzanm04}.
Using rigidly-rotating bars, \citet{elz_shl_02} computed the Liapunov
exponents of orbits and showed that chaos quickly dominates the
evolution when halos are triaxial and cuspy.  $N$-body simulations by
\citet{ber_etal_05} indeed show that bars are destroyed in cases where
the triaxiality in the potential is as small as $c/a \sim 0.9$ and the
halo is cuspy.  One way in which this fate can be avoided is for the
bar to alter the shape of the inner halo.  At present it is not clear
which systems can accomplish this and which cannot.  A better
understanding of when bars are destroyed in triaxial halos could
possibly lead to an important new constraint on the shapes and
profiles of dark matter halos.

\subsection{Bars within bars}
\label{ssec:s2bs}

While observations of largely gas-free early-type galaxies find an
abundance of nuclear bars within large scale bars (\citet{erw_spa_02}
found them in $\sim 30\%$ of barred S0-Sa galaxies), simulating them
has proved difficult.  Moreover, it is only recently that direct
observational evidence for kinematically decoupled primary and nuclear
bars in one system has been obtained \citep{cor_deb_agu_03}.
Therefore their dynamics have been poorly understood, in spite of the
fact that they have been postulated to be an important mechanism for
driving gas to small radii to feed supermassive black holes
\citep{shl_etal_89}.

Most numerical studies have required gas to form secondary bars
\citep{fri_mar_93, shl_hel_02, eng_shl_04}, but their presence in
gas-poor early-type galaxies suggests that gas is not the main
ingredient for forming secondary bars.  Stellar counter-rotation can
lead to counter-rotating bars \citep{sel_mer_94, friedl_96}.
\citet{rau_etal_02} were the first to succeed in producing
collisionless $N$-body simulations with both bars rotating in the same
sense.  Their secondary bars, which were vaguely spiral-like and
possibly hollow, rotated faster than the primary bars and survived for
several Gyrs.  Debattista \& Shen (2006, in preparation, see also Shen
\& Debattista in these proceedings) present further examples and
explore the mutual evolution of the two bars.  Now that $N$-body
simulations can achieve the high force and mass resolution needed to
form self-consistent nested bar systems it is hoped that progress in
understanding these systems will be more rapid than in the 30 years
since their discovery \citep{devauc_75}.


\section{Desiderata for the future}
\label{sec:conclusions}

The collisionless simulation of isolated galaxies is an endeavor over
thirty years old.  The subject is still rich, with several open
problems, and continues to be very active.  Algorithmically, if not
conceptually, it is the simplest problem that can be considered.
Various gravity solvers for its study are available (which are used
also in other areas of astronomy).  Despite much progress, the degree
of disagreement in the field, as described above, is surprising.  The
way to progress from this point is to compare directly the results of
different codes with each other, as has been done in other fields
\citep[e.g.][]{kang_etal_94, frenk_etal_99}.  Unfortunately the number
of such tests in galaxy evolution has been small.  Therefore a
detailed comparison between many different codes would be very
valuable at this time.  Ideally this would involve several
implementations of the same code type to establish behaviors in the
different types.  The actual tests to be performed should include
systems in which an analytic result is known (useful to establish
values of numerical parameters which are optimal) as well as systems
for which the result is not known in advance.  Furthermore, the
$N$-body tests of the type recently proposed by \citet{wei_kat_05}
provide a challenging and useful basis for assessing the effect of
noise on simulations.  Such a comparison will give the community
greater confidence that we are able to model correctly the most basic
level of galaxy evolution.


\acknowledgements V.P.D. is supported by a Brooks Prize Fellowship in
Astrophysics at the University of Washington.  I thank Tom Quinn,
Jerry Sellwood, Juntai Shen, Martin Weinberg, Peter Erwin and Kelly
Holley-Bockelmann for useful discussions, and the organizers of Bash
'05 for inviting me to this symposium.

\bibliographystyle{aj.bst}
\bibliography{allrefs}

\begin{thebibliography}{}

\bibitem[\protect\citeauthoryear{{Aarseth}}{{Aarseth}}{1963}]{aarset_63}
{Aarseth}, S.~J. 1963, \mnras, 126, 223

\bibitem[\protect\citeauthoryear{{Aguerri}, {Debattista}, \&
  {Corsini}}{{Aguerri} et~al.}{2003}]{agu_etal_03}
{Aguerri}, J.~A.~L., {Debattista}, V.~P.,  \& {Corsini}, E.~M. 2003, \mnras,
  338, 465

\bibitem[\protect\citeauthoryear{{Athanassoula}}{{Athanassoula}}{2002}]{athana%
_02}
{Athanassoula}, E. 2002, \apjl, 569, L83

\bibitem[\protect\citeauthoryear{{Athanassoula}}{{Athanassoula}}{2003}]{athana%
_03}
{Athanassoula}, E. 2003, \mnras, 341, 1179

\bibitem[\protect\citeauthoryear{{Athanassoula}, {Lambert}, \&
  {Dehnen}}{{Athanassoula} et~al.}{2005}]{ath_etal_05}
{Athanassoula}, E., {Lambert}, J.~C.,  \& {Dehnen}, W. 2005, \mnras, 363, 496

\bibitem[\protect\citeauthoryear{{Athanassoula} \& {Misiriotis}}{{Athanassoula}
  \& {Misiriotis}}{2002}]{ath_mis_02}
{Athanassoula}, E.,  \& {Misiriotis}, A. 2002, \mnras, 330, 35

\bibitem[\protect\citeauthoryear{{Avila-Reese} et~al.}{{Avila-Reese}
  et~al.}{2005}]{avi_etal_05}
{Avila-Reese}, V., {Carrillo}, A., {Valenzuela}, O.,  \& {Klypin}, A. 2005,
  \mnras, 361, 997

\bibitem[\protect\citeauthoryear{{Barnes} \& {Efstathiou}}{{Barnes} \&
  {Efstathiou}}{1987}]{bar_efs_87}
{Barnes}, J.,  \& {Efstathiou}, G. 1987, \apj, 319, 575

\bibitem[\protect\citeauthoryear{{Barnes} \& {Hut}}{{Barnes} \&
  {Hut}}{1986}]{bar_hut_86}
{Barnes}, J.,  \& {Hut}, P. 1986, \nat, 324, 446

\bibitem[\protect\citeauthoryear{{Barnes}, {Hut}, \& {Goodman}}{{Barnes}
  et~al.}{1986}]{barnes_etal_86}
{Barnes}, J., {Hut}, P.,  \& {Goodman}, J. 1986, \apj, 300, 112

\bibitem[\protect\citeauthoryear{{Berentzen} et~al.}{{Berentzen}
  et~al.}{1998}]{ber_etal_98}
{Berentzen}, I., {Heller}, C.~H., {Shlosman}, I.,  \& {Fricke}, K.~J. 1998,
  \mnras, 300, 49

\bibitem[\protect\citeauthoryear{{Berentzen}, {Shlosman}, \&
  {Jogee}}{{Berentzen} et~al.}{2005}]{ber_etal_05}
{Berentzen}, I., {Shlosman}, I.,  \& {Jogee}, S. 2005, astro-ph/0501352

\bibitem[\protect\citeauthoryear{{Bertin} et~al.}{{Bertin}
  et~al.}{1989a}]{ber_etal_89a}
{Bertin}, G., {Lin}, C.~C., {Lowe}, S.~A.,  \& {Thurstans}, R.~P. 1989a, \apj,
  338, 78

\bibitem[\protect\citeauthoryear{{Bertin} et~al.}{{Bertin}
  et~al.}{1989b}]{ber_etal_89b}
{Bertin}, G., {Lin}, C.~C., {Lowe}, S.~A.,  \& {Thurstans}, R.~P. 1989b, \apj,
  338, 104

\bibitem[\protect\citeauthoryear{{Bertschinger}}{{Bertschinger}}{1985}]{bertsc%
_85}
{Bertschinger}, E. 1985, \apjs, 58, 39

\bibitem[\protect\citeauthoryear{{Binney}, {Bissantz}, \& {Gerhard}}{{Binney}
  et~al.}{2000}]{bin_etal_00}
{Binney}, J., {Bissantz}, N.,  \& {Gerhard}, O. 2000, \apjl, 537, L99

\bibitem[\protect\citeauthoryear{{Binney}, {Gerhard}, \& {Silk}}{{Binney}
  et~al.}{2001}]{bin_etal_01}
{Binney}, J., {Gerhard}, O.,  \& {Silk}, J. 2001, \mnras, 321, 471

\bibitem[\protect\citeauthoryear{{Bissantz}, {Englmaier}, \&
  {Gerhard}}{{Bissantz} et~al.}{2003}]{bis_etal_03}
{Bissantz}, N., {Englmaier}, P.,  \& {Gerhard}, O. 2003, \mnras, 340, 949

\bibitem[\protect\citeauthoryear{{Blais-Ouellette}, {Amram}, \&
  {Carignan}}{{Blais-Ouellette} et~al.}{2001}]{bla_etal_01}
{Blais-Ouellette}, S., {Amram}, P.,  \& {Carignan}, C. 2001, \aj, 121, 1952

\bibitem[\protect\citeauthoryear{{Bode}, {Ostriker}, \& {Turok}}{{Bode}
  et~al.}{2001}]{bod_etal_01}
{Bode}, P., {Ostriker}, J.~P.,  \& {Turok}, N. 2001, \apj, 556, 93

\bibitem[\protect\citeauthoryear{{Bournaud} \& {Combes}}{{Bournaud} \&
  {Combes}}{2002}]{bou_com_02}
{Bournaud}, F.,  \& {Combes}, F. 2002, \aap, 392, 83

\bibitem[\protect\citeauthoryear{{Bournaud}, {Combes}, \& {Semelin}}{{Bournaud}
  et~al.}{2005}]{bou_etal_05}
{Bournaud}, F., {Combes}, F.,  \& {Semelin}, B. 2005, \mnras, 364, L18

\bibitem[\protect\citeauthoryear{{Bryan} \& {Norman}}{{Bryan} \&
  {Norman}}{1995}]{bry_nor_95}
{Bryan}, G.~L.,  \& {Norman}, M.~L. 1995, Bulletin of the American Astronomical
  Society, 27, 1421

\bibitem[\protect\citeauthoryear{{Bureau} \& {Athanassoula}}{{Bureau} \&
  {Athanassoula}}{1999}]{bur_ath_99}
{Bureau}, M.,  \& {Athanassoula}, E. 1999, \apj, 522, 686

\bibitem[\protect\citeauthoryear{{Bureau} \& {Athanassoula}}{{Bureau} \&
  {Athanassoula}}{2005}]{bur_ath_05}
{Bureau}, M.,  \& {Athanassoula}, E. 2005, \apj, 626, 159

\bibitem[\protect\citeauthoryear{{Bureau} \& {Freeman}}{{Bureau} \&
  {Freeman}}{1999}]{bur_fre_99}
{Bureau}, M.,  \& {Freeman}, K.~C. 1999, \aj, 118, 126

\bibitem[\protect\citeauthoryear{{Cen}}{{Cen}}{2001}]{cen_01}
{Cen}, R. 2001, \apjl, 546, L77

\bibitem[\protect\citeauthoryear{{Chung} \& {Bureau}}{{Chung} \&
  {Bureau}}{2004}]{chu_bur_04}
{Chung}, A.,  \& {Bureau}, M. 2004, \aj, 127, 3192

\bibitem[\protect\citeauthoryear{{Clutton-Brock}}{{Clutton-Brock}}{1972}]{clut%
to_72}
{Clutton-Brock}, M. 1972, \apss, 16, 101

\bibitem[\protect\citeauthoryear{{Col{\'{\i}}n}, {Avila-Reese}, \&
  {Valenzuela}}{{Col{\'{\i}}n} et~al.}{2000}]{col_etal_00}
{Col{\'{\i}}n}, P., {Avila-Reese}, V.,  \& {Valenzuela}, O. 2000, \apj, 542,
  622

\bibitem[\protect\citeauthoryear{{Combes} et~al.}{{Combes}
  et~al.}{1990}]{com_etal_90}
{Combes}, F., {Debbasch}, F., {Friedli}, D.,  \& {Pfenniger}, D. 1990, \aap,
  233, 82

\bibitem[\protect\citeauthoryear{{Combes} \& {Sanders}}{{Combes} \&
  {Sanders}}{1981}]{com_san_81}
{Combes}, F.,  \& {Sanders}, R.~H. 1981, \aap, 96, 164

\bibitem[\protect\citeauthoryear{{Contopoulos}}{{Contopoulos}}{1980}]{contop_8%
1}
{Contopoulos}, G. 1980, \aap, 81, 198

\bibitem[\protect\citeauthoryear{{Corsini}, {Debattista}, \&
  {Aguerri}}{{Corsini} et~al.}{2003}]{cor_deb_agu_03}
{Corsini}, E.~M., {Debattista}, V.~P.,  \& {Aguerri}, J.~A.~L. 2003, \apjl,
  599, L29

\bibitem[\protect\citeauthoryear{{Dav\'e}, {Dubinski}, \& {Hernquist}}{{Dav\'e}
  et~al.}{1997}]{dave_etal_97}
{Dav\'e}, R., {Dubinski}, J.,  \& {Hernquist}, L. 1997, New Astronomy, 2, 277

\bibitem[\protect\citeauthoryear{{de Blok}, {McGaugh}, \& {Rubin}}{{de Blok}
  et~al.}{2001}]{deblok_etal_01}
{de Blok}, W.~J.~G., {McGaugh}, S.~S.,  \& {Rubin}, V.~C. 2001, \aj, 122, 2396

\bibitem[\protect\citeauthoryear{{de Vaucouleurs}}{{de
  Vaucouleurs}}{1975}]{devauc_75}
{de Vaucouleurs}, G. 1975, \apjs, 29, 193

\bibitem[\protect\citeauthoryear{{Debattista}}{{Debattista}}{2003}]{debatt_03}
{Debattista}, V.~P. 2003, \mnras, 342, 1194

\bibitem[\protect\citeauthoryear{{Debattista} et~al.}{{Debattista}
  et~al.}{2004}]{deb_etal_04}
{Debattista}, V.~P., {Carollo}, C.~M., {Mayer}, L.,  \& {Moore}, B. 2004,
  \apjl, 604, L93

\bibitem[\protect\citeauthoryear{{Debattista} et~al.}{{Debattista}
  et~al.}{2005a}]{deb_etal_05a}
{Debattista}, V.~P., {Carollo}, C.~M., {Mayer}, L.,  \& {Moore}, B. 2005a,
  \apj, 628, 678

\bibitem[\protect\citeauthoryear{{Debattista}, {Corsini}, \&
  {Aguerri}}{{Debattista} et~al.}{2002}]{deb_cor_agu_02}
{Debattista}, V.~P., {Corsini}, E.~M.,  \& {Aguerri}, J.~A.~L. 2002, \mnras,
  332, 65

\bibitem[\protect\citeauthoryear{{Debattista} et~al.}{{Debattista}
  et~al.}{2005b}]{deb_etal_05b}
{Debattista}, V.~P., {Mayer}, L., {Carollo}, C.~M., {Moore}, B., {Wadsley}, J.,
   \& {Quinn}, T. 2005b, astro-ph/0509310

\bibitem[\protect\citeauthoryear{{Debattista} \& {Sellwood}}{{Debattista} \&
  {Sellwood}}{1998}]{deb_sel_98}
{Debattista}, V.~P.,  \& {Sellwood}, J.~A. 1998, \apjl, 493, L5

\bibitem[\protect\citeauthoryear{{Debattista} \& {Sellwood}}{{Debattista} \&
  {Sellwood}}{2000}]{deb_sel_00}
{Debattista}, V.~P.,  \& {Sellwood}, J.~A. 2000, \apj, 543, 704

\bibitem[\protect\citeauthoryear{{Debattista} \& {Williams}}{{Debattista} \&
  {Williams}}{2004}]{deb_wil_04}
{Debattista}, V.~P.,  \& {Williams}, T.~B. 2004, \apj, 605, 714

\bibitem[\protect\citeauthoryear{{Dehnen}}{{Dehnen}}{2000}]{dehnen_00}
{Dehnen}, W. 2000, \apjl, 536, L39

\bibitem[\protect\citeauthoryear{{Dubinski}}{{Dubinski}}{1994}]{dubins_94}
{Dubinski}, J. 1994, \apj, 431, 617

\bibitem[\protect\citeauthoryear{{Dubinski}}{{Dubinski}}{1996}]{dubins_96}
{Dubinski}, J. 1996, New Astronomy, 1, 133

\bibitem[\protect\citeauthoryear{{Dubinski} \& {Carlberg}}{{Dubinski} \&
  {Carlberg}}{1991}]{dub_car_91}
{Dubinski}, J.,  \& {Carlberg}, R.~G. 1991, \apj, 378, 496

\bibitem[\protect\citeauthoryear{{Earn} \& {Sellwood}}{{Earn} \&
  {Sellwood}}{1995}]{ear_sel_95}
{Earn}, D.~J.~D.,  \& {Sellwood}, J.~A. 1995, \apj, 451, 533

\bibitem[\protect\citeauthoryear{{El-Zant} \& {Shlosman}}{{El-Zant} \&
  {Shlosman}}{2002}]{elz_shl_02}
{El-Zant}, A.,  \& {Shlosman}, I. 2002, \apj, 577, 626

\bibitem[\protect\citeauthoryear{{Englmaier} \& {Shlosman}}{{Englmaier} \&
  {Shlosman}}{2004}]{eng_shl_04}
{Englmaier}, P.,  \& {Shlosman}, I. 2004, \apjl, 617, L115

\bibitem[\protect\citeauthoryear{{Erwin}}{{Erwin}}{2005}]{erwin_05}
{Erwin}, P. 2005, \mnras, 364, 283

\bibitem[\protect\citeauthoryear{{Erwin} \& {Sparke}}{{Erwin} \&
  {Sparke}}{2002}]{erw_spa_02}
{Erwin}, P.,  \& {Sparke}, L.~S. 2002, \aj, 124, 65

\bibitem[\protect\citeauthoryear{{Eskridge} et~al.}{{Eskridge}
  et~al.}{2000}]{esk_etal_00}
{Eskridge}, P.~B., et~al. 2000, \aj, 119, 536

\bibitem[\protect\citeauthoryear{{Evans} \& {Read}}{{Evans} \&
  {Read}}{1998a}]{eva_rea_98a}
{Evans}, N.~W.,  \& {Read}, J.~C.~A. 1998a, \mnras, 300, 83

\bibitem[\protect\citeauthoryear{{Evans} \& {Read}}{{Evans} \&
  {Read}}{1998b}]{eva_rea_98b}
{Evans}, N.~W.,  \& {Read}, J.~C.~A. 1998b, \mnras, 300, 106

\bibitem[\protect\citeauthoryear{{Fillmore} \& {Goldreich}}{{Fillmore} \&
  {Goldreich}}{1984}]{fil_gol_84}
{Fillmore}, J.~A.,  \& {Goldreich}, P. 1984, \apj, 281, 1

\bibitem[\protect\citeauthoryear{{Frenk} et~al.}{{Frenk}
  et~al.}{1999}]{frenk_etal_99}
{Frenk}, C.~S., et~al. 1999, \apj, 525, 554

\bibitem[\protect\citeauthoryear{{Frenk} et~al.}{{Frenk}
  et~al.}{1988}]{frenk_etal_88}
{Frenk}, C.~S., {White}, S.~D.~M., {Davis}, M.,  \& {Efstathiou}, G. 1988,
  \apj, 327, 507

\bibitem[\protect\citeauthoryear{{Friedli}}{{Friedli}}{1996}]{friedl_96}
{Friedli}, D. 1996, \aap, 312, 761

\bibitem[\protect\citeauthoryear{{Friedli} \& {Benz}}{{Friedli} \&
  {Benz}}{1993}]{fri_ben_93}
{Friedli}, D.,  \& {Benz}, W. 1993, \aap, 268, 65

\bibitem[\protect\citeauthoryear{{Friedli} \& {Martinet}}{{Friedli} \&
  {Martinet}}{1993}]{fri_mar_93}
{Friedli}, D.,  \& {Martinet}, L. 1993, \aap, 277, 27

\bibitem[\protect\citeauthoryear{{Fux}}{{Fux}}{1999}]{fux_99}
{Fux}, R. 1999, \aap, 345, 787

\bibitem[\protect\citeauthoryear{{Gerhard}}{{Gerhard}}{1993}]{gerhar_93}
{Gerhard}, O.~E. 1993, \mnras, 265, 213

\bibitem[\protect\citeauthoryear{{Gerhard} \& {Binney}}{{Gerhard} \&
  {Binney}}{1985}]{ger_bin_85}
{Gerhard}, O.~E.,  \& {Binney}, J. 1985, \mnras, 216, 467

\bibitem[\protect\citeauthoryear{{Gerssen}, {Kuijken}, \&
  {Merrifield}}{{Gerssen} et~al.}{1999}]{gersse_etal_99}
{Gerssen}, J., {Kuijken}, K.,  \& {Merrifield}, M.~R. 1999, \mnras, 306, 926

\bibitem[\protect\citeauthoryear{{Gerssen}, {Kuijken}, \&
  {Merrifield}}{{Gerssen} et~al.}{2003}]{gersse_etal_03}
{Gerssen}, J., {Kuijken}, K.,  \& {Merrifield}, M.~R. 2003, \mnras, 345, 261

\bibitem[\protect\citeauthoryear{{Gnedin} \& {Zhao}}{{Gnedin} \&
  {Zhao}}{2002}]{gne_zha_02}
{Gnedin}, O.~Y.,  \& {Zhao}, H. 2002, \mnras, 333, 299

\bibitem[\protect\citeauthoryear{{Goodman}}{{Goodman}}{2000}]{goodma_00}
{Goodman}, J. 2000, New Astronomy, 5, 103

\bibitem[\protect\citeauthoryear{{Hasan} \& {Norman}}{{Hasan} \&
  {Norman}}{1990}]{has_nor_90}
{Hasan}, H.,  \& {Norman}, C. 1990, \apj, 361, 69

\bibitem[\protect\citeauthoryear{{Hasan}, {Pfenniger}, \& {Norman}}{{Hasan}
  et~al.}{1993}]{has_etal_93}
{Hasan}, H., {Pfenniger}, D.,  \& {Norman}, C. 1993, \apj, 409, 91

\bibitem[\protect\citeauthoryear{{Hernquist}}{{Hernquist}}{1987}]{hernqu_87}
{Hernquist}, L. 1987, \apjs, 64, 715

\bibitem[\protect\citeauthoryear{{Hernquist}}{{Hernquist}}{1990}]{hernqu_90}
{Hernquist}, L. 1990, \apj, 356, 359

\bibitem[\protect\citeauthoryear{{Hernquist} \& {Barnes}}{{Hernquist} \&
  {Barnes}}{1990}]{her_bar_90}
{Hernquist}, L.,  \& {Barnes}, J.~E. 1990, \apj, 349, 562

\bibitem[\protect\citeauthoryear{{Hernquist} \& {Ostriker}}{{Hernquist} \&
  {Ostriker}}{1992}]{her_ost_92}
{Hernquist}, L.,  \& {Ostriker}, J.~P. 1992, \apj, 386, 375

\bibitem[\protect\citeauthoryear{{Hernquist} \& {Weinberg}}{{Hernquist} \&
  {Weinberg}}{1992}]{her_wei_92}
{Hernquist}, L.,  \& {Weinberg}, M.~D. 1992, \apj, 400, 80

\bibitem[\protect\citeauthoryear{{Hogan} \& {Dalcanton}}{{Hogan} \&
  {Dalcanton}}{2000}]{hog_dal_00}
{Hogan}, C.~J.,  \& {Dalcanton}, J.~J. 2000, \prd, 62, 063511

\bibitem[\protect\citeauthoryear{{Hohl}}{{Hohl}}{1971}]{hohl_71}
{Hohl}, F. 1971, \apj, 168, 343

\bibitem[\protect\citeauthoryear{{Hohl}}{{Hohl}}{1973}]{hohl_73}
{Hohl}, F. 1973, \apj, 184, 353

\bibitem[\protect\citeauthoryear{{Holley-Bockelmann}, {Weinberg}, \&
  {Katz}}{{Holley-Bockelmann} et~al.}{2005}]{hol_etal_05}
{Holley-Bockelmann}, K., {Weinberg}, M.,  \& {Katz}, N. 2005, \mnras, 363, 991

\bibitem[\protect\citeauthoryear{{Hozumi} \& {Hernquist}}{{Hozumi} \&
  {Hernquist}}{2005}]{hoz_her_05}
{Hozumi}, S.,  \& {Hernquist}, L. 2005, \pasj, 57, 719

\bibitem[\protect\citeauthoryear{{Inagaki}, {Nishida}, \& {Sellwood}}{{Inagaki}
  et~al.}{1984}]{ina_etal_84}
{Inagaki}, S., {Nishida}, M.~T.,  \& {Sellwood}, J.~A. 1984, \mnras, 210, 589

\bibitem[\protect\citeauthoryear{{Jing} \& {Suto}}{{Jing} \&
  {Suto}}{2000}]{jin_sut_00}
{Jing}, Y.~P.,  \& {Suto}, Y. 2000, \apjl, 529, L69

\bibitem[\protect\citeauthoryear{{Jing} \& {Suto}}{{Jing} \&
  {Suto}}{2002}]{jin_sut_02}
{Jing}, Y.~P.,  \& {Suto}, Y. 2002, \apj, 574, 538

\bibitem[\protect\citeauthoryear{{Kalnajs}}{{Kalnajs}}{1972}]{kalnaj_72}
{Kalnajs}, A.~J. 1972, \apj, 175, 63

\bibitem[\protect\citeauthoryear{{Kalnajs}}{{Kalnajs}}{1978}]{kalnaj_78}
{Kalnajs}, A.~J. 1978, in IAU Symp. 77: Structure and Properties of Nearby
  Galaxies, 113

\bibitem[\protect\citeauthoryear{{Kang} et~al.}{{Kang}
  et~al.}{1994}]{kang_etal_94}
{Kang}, H., {Ostriker}, J.~P., {Cen}, R., {Ryu}, D., {Hernquist}, L., {Evrard},
  A.~E., {Bryan}, G.~L.,  \& {Norman}, M.~L. 1994, \apj, 430, 83

\bibitem[\protect\citeauthoryear{{Kaplinghat}, {Knox}, \&
  {Turner}}{{Kaplinghat} et~al.}{2000}]{kap_etal_00}
{Kaplinghat}, M., {Knox}, L.,  \& {Turner}, M.~S. 2000, Physical Review
  Letters, 85, 3335

\bibitem[\protect\citeauthoryear{{Kazantzidis} et~al.}{{Kazantzidis}
  et~al.}{2004}]{kkzanm04}
{Kazantzidis}, S., {Kravtsov}, A.~V., {Zentner}, A.~R., {Allgood}, B., {Nagai},
  D.,  \& {Moore}, B. 2004, \apjl, 611, L73

\bibitem[\protect\citeauthoryear{{Knapen}}{{Knapen}}{1999}]{knapen_99}
{Knapen}, J.~H. 1999, in ASP Conf. Ser. 187: The Evolution of Galaxies on
  Cosmological Timescales, 72

\bibitem[\protect\citeauthoryear{{Kormendy} \& {Kennicutt}}{{Kormendy} \&
  {Kennicutt}}{2004}]{kor_ken_04}
{Kormendy}, J.,  \& {Kennicutt}, R.~C. 2004, \araa, 42, 603

\bibitem[\protect\citeauthoryear{{Kravtsov}, {Klypin}, \&
  {Khokhlov}}{{Kravtsov} et~al.}{1997}]{kra_etal_97}
{Kravtsov}, A.~V., {Klypin}, A.~A.,  \& {Khokhlov}, A.~M. 1997, \apjs, 111, 73

\bibitem[\protect\citeauthoryear{{Kuijken} \& {Merrifield}}{{Kuijken} \&
  {Merrifield}}{1995}]{kui_mer_95}
{Kuijken}, K.,  \& {Merrifield}, M.~R. 1995, \apjl, 443, L13

\bibitem[\protect\citeauthoryear{{Lin} \& {Shu}}{{Lin} \&
  {Shu}}{1964}]{lin_shu_64}
{Lin}, C.~C.,  \& {Shu}, F.~H. 1964, \apj, 140, 646

\bibitem[\protect\citeauthoryear{{Lindblad} \& {Kristen}}{{Lindblad} \&
  {Kristen}}{1996}]{lin_kri_96}
{Lindblad}, P.~A.~B.,  \& {Kristen}, H. 1996, \aap, 313, 733

\bibitem[\protect\citeauthoryear{{Lindblad}, {Lindblad}, \&
  {Athanassoula}}{{Lindblad} et~al.}{1996}]{lin_etal_96}
{Lindblad}, P.~A.~B., {Lindblad}, P.~O.,  \& {Athanassoula}, E. 1996, \aap,
  313, 65

\bibitem[\protect\citeauthoryear{{Little} \& {Carlberg}}{{Little} \&
  {Carlberg}}{1991}]{lit_car_91}
{Little}, B.,  \& {Carlberg}, R.~G. 1991, \mnras, 250, 161

\bibitem[\protect\citeauthoryear{{L{\"u}tticke}, {Dettmar}, \&
  {Pohlen}}{{L{\"u}tticke} et~al.}{2000}]{lut_etal_00}
{L{\"u}tticke}, R., {Dettmar}, R.-J.,  \& {Pohlen}, M. 2000, \aaps, 145, 405

\bibitem[\protect\citeauthoryear{{Martinez-Valpuesta}, {Shlosman}, \&
  {Heller}}{{Martinez-Valpuesta} et~al.}{2005}]{mar_etal_05}
{Martinez-Valpuesta}, I., {Shlosman}, I.,  \& {Heller}, C. 2005,
  astro-ph/0507219

\bibitem[\protect\citeauthoryear{{Masset} \& {Tagger}}{{Masset} \&
  {Tagger}}{1997}]{mas_tag_97}
{Masset}, F.,  \& {Tagger}, M. 1997, \aap, 322, 442

\bibitem[\protect\citeauthoryear{{Matthews} \& {Gallagher}}{{Matthews} \&
  {Gallagher}}{2002}]{mat_gal_02}
{Matthews}, L.~D.,  \& {Gallagher}, J.~S. 2002, \apjs, 141, 429

\bibitem[\protect\citeauthoryear{{McMillan} \& {Dehnen}}{{McMillan} \&
  {Dehnen}}{2005}]{mcm_deh_05}
{McMillan}, P.~J.,  \& {Dehnen}, W. 2005, \mnras, 363, 1205

\bibitem[\protect\citeauthoryear{{Merrifield} \& {Kuijken}}{{Merrifield} \&
  {Kuijken}}{1995}]{mer_kui_95}
{Merrifield}, M.~R.,  \& {Kuijken}, K. 1995, \mnras, 274, 933

\bibitem[\protect\citeauthoryear{{Merrifield} \& {Kuijken}}{{Merrifield} \&
  {Kuijken}}{1999}]{mer_kui_99}
{Merrifield}, M.~R.,  \& {Kuijken}, K. 1999, \aap, 345, L47

\bibitem[\protect\citeauthoryear{{Miller}}{{Miller}}{1976}]{miller_76}
{Miller}, R.~H. 1976, Journal of Computational Physics, 21, 400

\bibitem[\protect\citeauthoryear{{Miller} \& {Prendergast}}{{Miller} \&
  {Prendergast}}{1968}]{mil_pre_68}
{Miller}, R.~H.,  \& {Prendergast}, K.~H. 1968, \apj, 151, 699

\bibitem[\protect\citeauthoryear{{Miller}, {Prendergast}, \& {Quirk}}{{Miller}
  et~al.}{1970}]{mil_etal_70}
{Miller}, R.~H., {Prendergast}, K.~H.,  \& {Quirk}, W.~J. 1970, \apj, 161, 903

\bibitem[\protect\citeauthoryear{{Moore} et~al.}{{Moore}
  et~al.}{1998}]{moo_etal_98}
{Moore}, B., {Governato}, F., {Quinn}, T., {Stadel}, J.,  \& {Lake}, G. 1998,
  \apjl, 499, L5

\bibitem[\protect\citeauthoryear{{Navarro}, {Eke}, \& {Frenk}}{{Navarro}
  et~al.}{1996}]{nav_etal_96}
{Navarro}, J.~F., {Eke}, V.~R.,  \& {Frenk}, C.~S. 1996, \mnras, 283, L72

\bibitem[\protect\citeauthoryear{{Navarro}, {Frenk}, \& {White}}{{Navarro}
  et~al.}{1997}]{nav_etal_97_nfw}
{Navarro}, J.~F., {Frenk}, C.~S.,  \& {White}, S.~D.~M. 1997, \apj, 490, 493

\bibitem[\protect\citeauthoryear{{Nishida} et~al.}{{Nishida}
  et~al.}{1981}]{nis_etal_81}
{Nishida}, M.~T., {Yoshizawa}, M., {Watanabe}, Y., {Inagaki}, S.,  \& {Kato},
  S. 1981, \pasj, 33, 567

\bibitem[\protect\citeauthoryear{{Norman}, {Sellwood}, \& {Hasan}}{{Norman}
  et~al.}{1996}]{nor_etal_96}
{Norman}, C.~A., {Sellwood}, J.~A.,  \& {Hasan}, H. 1996, \apj, 462, 114

\bibitem[\protect\citeauthoryear{{O'Neill} \& {Dubinski}}{{O'Neill} \&
  {Dubinski}}{2003}]{one_dub_03}
{O'Neill}, J.~K.,  \& {Dubinski}, J. 2003, \mnras, 346, 251

\bibitem[\protect\citeauthoryear{{Peebles}}{{Peebles}}{2000}]{peeble_00}
{Peebles}, P.~J.~E. 2000, \apjl, 534, L127

\bibitem[\protect\citeauthoryear{{P{\'e}rez}, {Fux}, \& {Freeman}}{{P{\'e}rez}
  et~al.}{2004}]{per_etal_04}
{P{\'e}rez}, I., {Fux}, R.,  \& {Freeman}, K. 2004, \aap, 424, 799

\bibitem[\protect\citeauthoryear{{Pfenniger}}{{Pfenniger}}{1984}]{pfenni_84}
{Pfenniger}, D. 1984, \aap, 134, 373

\bibitem[\protect\citeauthoryear{{Pfenniger} \& {Friedli}}{{Pfenniger} \&
  {Friedli}}{1991}]{pfe_fri_91}
{Pfenniger}, D.,  \& {Friedli}, D. 1991, \aap, 252, 75

\bibitem[\protect\citeauthoryear{{Pohlen}}{{Pohlen}}{2002}]{pohlen_phd}
{Pohlen}, M. 2002, Ph.D.~Thesis

\bibitem[\protect\citeauthoryear{{Power} et~al.}{{Power}
  et~al.}{2003}]{pow_etal_03}
{Power}, C., {Navarro}, J.~F., {Jenkins}, A., {Frenk}, C.~S., {White},
  S.~D.~M., {Springel}, V., {Stadel}, J.,  \& {Quinn}, T. 2003, \mnras, 338, 14

\bibitem[\protect\citeauthoryear{{Raha} et~al.}{{Raha}
  et~al.}{1991}]{rah_etal_91}
{Raha}, N., {Sellwood}, J.~A., {James}, R.~A.,  \& {Kahn}, F.~D. 1991, \nat,
  352, 411

\bibitem[\protect\citeauthoryear{{Rautiainen} \& {Salo}}{{Rautiainen} \&
  {Salo}}{1999}]{rau_sal_99}
{Rautiainen}, P.,  \& {Salo}, H. 1999, \aap, 348, 737

\bibitem[\protect\citeauthoryear{{Rautiainen}, {Salo}, \&
  {Laurikainen}}{{Rautiainen} et~al.}{2002}]{rau_etal_02}
{Rautiainen}, P., {Salo}, H.,  \& {Laurikainen}, E. 2002, \mnras, 337, 1233

\bibitem[\protect\citeauthoryear{{Sellwood}}{{Sellwood}}{1980}]{sellwo_80}
{Sellwood}, J.~A. 1980, \aap, 89, 296

\bibitem[\protect\citeauthoryear{{Sellwood}}{{Sellwood}}{1997}]{sellwo_97}
{Sellwood}, J.~A. 1997, in ASP Conf. Ser. 123: Computational Astrophysics; 12th
  Kingston Meeting on Theoretical Astrophysics, 215

\bibitem[\protect\citeauthoryear{{Sellwood}}{{Sellwood}}{2000}]{sellwo_00b}
{Sellwood}, J.~A. 2000, \apss, 272, 31

\bibitem[\protect\citeauthoryear{{Sellwood}}{{Sellwood}}{2002}]{sellwo_02}
{Sellwood}, J.~A. 2002, in The shapes of galaxies and their dark halos,
  Proceedings of the Yale Cosmology Workshop "The Shapes of Galaxies and Their
  Dark Matter Halos", New Haven, Connecticut, USA, 28-30 May 2001. Edited by
  Priyamvada Natarajan. Singapore: World Scientific, 2002, ISBN 9810248482,
  p.123, 123

\bibitem[\protect\citeauthoryear{{Sellwood}}{{Sellwood}}{2003}]{sellwo_03}
{Sellwood}, J.~A. 2003, \apj, 587, 638

\bibitem[\protect\citeauthoryear{{Sellwood}}{{Sellwood}}{2005}]{sellwo_05}
{Sellwood}, J.~A. 2005, astro-ph/0407533

\bibitem[\protect\citeauthoryear{{Sellwood} \& {Binney}}{{Sellwood} \&
  {Binney}}{2002}]{sel_bin_02}
{Sellwood}, J.~A.,  \& {Binney}, J.~J. 2002, \mnras, 336, 785

\bibitem[\protect\citeauthoryear{{Sellwood} \& {Debattista}}{{Sellwood} \&
  {Debattista}}{2005}]{sel_deb_05}
{Sellwood}, J.~A.,  \& {Debattista}, V.~P. 2005, astro-ph/0511155

\bibitem[\protect\citeauthoryear{{Sellwood} \& {Evans}}{{Sellwood} \&
  {Evans}}{2001}]{sel_eva_01}
{Sellwood}, J.~A.,  \& {Evans}, N.~W. 2001, \apj, 546, 176

\bibitem[\protect\citeauthoryear{{Sellwood} \& {Kahn}}{{Sellwood} \&
  {Kahn}}{1991}]{sel_kah_91}
{Sellwood}, J.~A.,  \& {Kahn}, F.~D. 1991, \mnras, 250, 278

\bibitem[\protect\citeauthoryear{{Sellwood} \& {Lin}}{{Sellwood} \&
  {Lin}}{1989}]{sel_lin_89}
{Sellwood}, J.~A.,  \& {Lin}, D.~N.~C. 1989, \mnras, 240, 991

\bibitem[\protect\citeauthoryear{{Sellwood} \& {Merritt}}{{Sellwood} \&
  {Merritt}}{1994}]{sel_mer_94}
{Sellwood}, J.~A.,  \& {Merritt}, D. 1994, \apj, 425, 530

\bibitem[\protect\citeauthoryear{{Shen} \& {Sellwood}}{{Shen} \&
  {Sellwood}}{2004}]{she_sel_04}
{Shen}, J.,  \& {Sellwood}, J.~A. 2004, \apj, 604, 614

\bibitem[\protect\citeauthoryear{{Shlosman}, {Frank}, \& {Begelman}}{{Shlosman}
  et~al.}{1989}]{shl_etal_89}
{Shlosman}, I., {Frank}, J.,  \& {Begelman}, M.~C. 1989, \nat, 338, 45

\bibitem[\protect\citeauthoryear{{Shlosman} \& {Heller}}{{Shlosman} \&
  {Heller}}{2002}]{shl_hel_02}
{Shlosman}, I.,  \& {Heller}, C.~H. 2002, \apj, 565, 921

\bibitem[\protect\citeauthoryear{{Spergel} \& {Steinhardt}}{{Spergel} \&
  {Steinhardt}}{2000}]{spe_ste_00}
{Spergel}, D.~N.,  \& {Steinhardt}, P.~J. 2000, Physical Review Letters, 84,
  3760

\bibitem[\protect\citeauthoryear{{Springel}, {Yoshida}, \& {White}}{{Springel}
  et~al.}{2001}]{spr_etal_01}
{Springel}, V., {Yoshida}, N.,  \& {White}, S.~D.~M. 2001, New Astronomy, 6, 79

\bibitem[\protect\citeauthoryear{{Stadel}}{{Stadel}}{2001}]{stadel_phd}
{Stadel}, J.~G. 2001, Ph.D.~Thesis

\bibitem[\protect\citeauthoryear{{Toomre}}{{Toomre}}{1981}]{toomre_81}
{Toomre}, A. 1981, in Structure and Evolution of Normal Galaxies, ed. S.~M.,
  Fall \& D. Lynden-Bell (Cambridge: Cambridge University Press), 111

\bibitem[\protect\citeauthoryear{{Tremaine} \& {Weinberg}}{{Tremaine} \&
  {Weinberg}}{1984}]{tre_wei_84}
{Tremaine}, S.,  \& {Weinberg}, M.~D. 1984, \apjl, 282, L5

\bibitem[\protect\citeauthoryear{{Valenzuela} \& {Klypin}}{{Valenzuela} \&
  {Klypin}}{2003}]{val_kly_03}
{Valenzuela}, O.,  \& {Klypin}, A. 2003, \mnras, 345, 406

\bibitem[\protect\citeauthoryear{{van der Marel} \& {Franx}}{{van der Marel} \&
  {Franx}}{1993}]{vdm_fra_93}
{van der Marel}, R.~P.,  \& {Franx}, M. 1993, \apj, 407, 525

\bibitem[\protect\citeauthoryear{{Watanabe} et~al.}{{Watanabe}
  et~al.}{1981}]{wat_etal_81}
{Watanabe}, Y., {Inagaki}, S., {Nishida}, M.~T., {Tanaka}, Y.~D.,  \& {Kato},
  S. 1981, \pasj, 33, 541

\bibitem[\protect\citeauthoryear{{Weinberg}}{{Weinberg}}{1985}]{weinbe_85}
{Weinberg}, M.~D. 1985, \mnras, 213, 451

\bibitem[\protect\citeauthoryear{{Weinberg}}{{Weinberg}}{1996}]{weinbe_96}
{Weinberg}, M.~D. 1996, \apj, 470, 715

\bibitem[\protect\citeauthoryear{{Weinberg}}{{Weinberg}}{2005}]{weinbe_04}
{Weinberg}, M.~D. 2005, astro-ph/0404169

\bibitem[\protect\citeauthoryear{{Weinberg} \& {Katz}}{{Weinberg} \&
  {Katz}}{2002}]{wei_kat_02}
{Weinberg}, M.~D.,  \& {Katz}, N. 2002, \apj, 580, 627

\bibitem[\protect\citeauthoryear{{Weinberg} \& {Katz}}{{Weinberg} \&
  {Katz}}{2005}]{wei_kat_05}
{Weinberg}, M.~D.,  \& {Katz}, N. 2005, astro-ph/0508166

\bibitem[\protect\citeauthoryear{{Weiner}, {Sellwood}, \& {Williams}}{{Weiner}
  et~al.}{2001}]{wei_etal_01b}
{Weiner}, B.~J., {Sellwood}, J.~A.,  \& {Williams}, T.~B. 2001, \apj, 546, 931

\bibitem[\protect\citeauthoryear{{Zang}}{{Zang}}{1976}]{zang_phd}
{Zang}, T.~A. 1976, Ph.D.~Thesis

\bibitem[\protect\citeauthoryear{{Zel'Dovich}}{{Zel'Dovich}}{1970}]{zeldov_70}
{Zel'Dovich}, Y.~B. 1970, \aap, 5, 84

\end{thebibliography}

\end{document}